# Multiplexed Extrinsic Fiber Fabry-Perot Interferometric Sensors: Resolution Limits

Nikolai A. Ushakov, Leonid B. Liokumovich

*Abstract*— Resolution investigation of multiplexed displacement sensors based on extrinsic fiber Fabry-Perot interferometers (EFPI) was carried out. The cases of serial and parallel configurations were considered, the analysis of the problems and the advantages of the both was performed. An analytical model, describing the resolution limits for the both configurations was developed. Serial and parallel multiplexing schemes have been experimentally implemented with 3 and 4 sensing elements, respectively. For the both configurations the achieved optical path difference (OPD) standard deviations were between 30 and 80 pm, which is, to the best of our knowledge, more than an order better than any other multiplexed EFPI resolution ever reported. A good correspondence between experimental results and analytical predictions was demonstrated. A mathematical apparatus, relating the attained sensors resolutions with the number of multiplexed sensors and the optical setup parameters was developed, also demonstrating good correspondence with the experimental results. The main origins of possible sensors cross-talk are described, with the supporting experiments, proving an importance of its consideration. Some recommendations, based on the theoretical analysis and experimental studies, dedicated to optimization of sensors resolution and elimination of the cross-talk influence are provided.

## I. Introduction

During the last two decades a great progress in manufacture and implementation of the fiber optic sensors based on the extrinsic Fabry-Perot interferometers (EFPI) [1] has been achieved by the academic institutions and commercial companies. Such sensors demonstrate a great dynamic measurement range and a high resolution [2], [3]. Sensors of a great variety of physical quantities have been designed and implemented. The most commonly used EFPI OPD demodulation approaches are the white-light interferometry, using a scanning readout interferometer [4] and wavelength-domain interferometry, in which measurement and further processing of the interrogated interferometer spectral function is used to find its OPD [5]–[7]. Both these techniques provide an ability to obtain the absolute OPD value and to track the signals of a system of multiplexed interferometers with different OPD values [8], [9]. However, the spectrum measurement approach is more advantageous, with the best OPD resolution of a single sensor about 10 picometers [2], [3]. However, the best achieved resolution for the multiplexed sensors was about 1 nanometer [9]. Also, despite the experimental progress in the field, the theory of such sensors isn't well developed and, to the best of our knowledge, no analytical description of their possible performance is presented in the literature. So, the scope of the current work is to introduce some fundamentals of such description and to achieve the OPD resolution comparable with the one attained for single-sensor devices. This paper delivers an expansion of the theoretical analysis and experimental work, presented on an OFS23 conference earlier [10]. The analysis of the noise sources and the light propagation inside the EFPI cavity will be based on an earlier paper [11], dedicated to the resolution limits analysis of single EFPI sensors.

Throughout this paper we consider the case of registering the spectrum of the light reflected from the sensor, which is the most common case for both single-sensor and multiplexed systems. The spectral function of a low-finesse Fabry-Perot interferometer contains an oscillating quasi-harmonic component $S_M \cdot S(L, \lambda)$, which in this case is given by [11]:

$$S(L, \lambda) = \cos[4\pi n L/\lambda + \gamma_R(L, \lambda)], \quad (1)$$

$$S_M = 2\sqrt{R_1 R_2 \cdot \eta(L)},\ \eta(L) = \left(\pi n w_0^2\right)^2 \Big/ \left[ L^2 \lambda^2 + \left(\pi n w_0^2\right)^2 \right], \quad (2)$$

$$\gamma_R(L, \lambda) = \psi_R + \varphi_R = \operatorname{atan}\left(4 L^3 / z_R^3 + 3 L / z_R\right) + \varphi_R \quad (3)$$

where a Gaussian beam assumption was applied to the fiber mode and the beam inside the cavity; $L$ is EFPI cavity length; $R_1$ and $R_2$ are the mirrors reflectivities; $\eta(L)$ is a coupling coefficient of a light beam, irradiated by a fiber mode, travelled twice the cavity length distance and captured by the fiber mode [11]; $w_0$ is an effective radius of the mode at the output of the first fiber; $\lambda$ is the light wavelength; $n$ is the refractive index of the media between the mirrors; $z_R = \pi n w_0^2/\lambda$ – Rayleigh length of the intra-cavity Gaussian beam; the argument additive $\gamma_R(L, \lambda)$ contains a phase shift $\psi_R$, induced by the light diffraction inside the cavity, and a phase $\varphi_R$, induced by the mirrors (typically for dielectric mirrors $\varphi_R = \pi$). Examples of EFPI cavities are schematically shown in fig. 1.

The work was supported by the program State Tasks for Higher Educational Institutions (projects no. 3.1446.2014K and 2014/184).

The authors are with the St. Petersburg State Polytechnical University, Department of Radiophysics, St. Petersburg, 195251, Russia (e-mail: n.ushakoff@spbstu.ru, leonid@spbstu.ru).

Eq. (1) gives a quasi-harmonic oscillation with respect to the wavelength shift ($\lambda - \lambda_0$) with frequency $4\pi nL/\lambda_0^2$, ($\lambda_0$ is a mean wavelength of the spectra measurement interval).

## II. Multiplexed sensors features

Transfer function of a system of $N$ multiplexed EFPIs with different OPDs $L_j$, $j=1..N$ is a superposition of oscillations (1) with different frequencies and can be expressed as

$$S_{\text{MULT}}(L_1,\ldots,L_N,\lambda) = \sum_{j=1}^{N} S'_{Mj} S_j(\lambda) + H(\lambda), \quad (4)$$

where $S_j(\lambda)=S(L_j, \lambda)$; $S'_{Mj}$ are the amplitudes of each sensor responce; $H(\lambda)$ contains quasi-static component, additive noises and a number of parasitic components of a form (1) with equivalent OPDs different from $L_1,\ldots, L_N$. Since (4) is a superposition of $S_j(\lambda)$ with different frequencies $4\pi n_j L_j/\lambda_0^2$, applying band pass filters to the signal $S_{\text{mult}}$, extracting partial components $S_j(\lambda)$ and applying the OPD demodulation approaches developed for a single sensor, one can obtain the readings $L_{rj}$ from each multiplexed sensor.

Considering the multiplexing possibilities, one of the first questions is the maximal possible number of sensors $N_{\max}$ that can be interrogated by a given setup. A possibility to distinguish components $S_j(\lambda)$ with different OPDs is related with the filtering resolution, given by

$$\Delta L = \lambda_0^2 / 2n\Lambda, \quad (5)$$

where $\Lambda$ is the width of the wavelength scanning interval. $\Delta L$ will determine the minimal feasible OPD value $L_j > \Delta L$ and minimal OPDs difference $|L_j - L_k| \gg \Delta L$. On the other hand, maximal OPD value $L_{\max}$ will be limited by the spectral resolution $\Delta'$ for initially registered $S_{\text{mult}}(\lambda)$ (determined by spectral width of a tunable light source or by resolution of a spectrum analyzer). This limitation can be written as $L_{\max} \ll \lambda_0^2/4\pi n\Delta'$, otherwise the component $S(L_{\max}, \lambda)$ will not be registered correctly. Assuming typical values $\Delta' \sim 10$ pm, $\Delta L \sim 10$ μm, one obtains maximal number of sensors about 100. However, as will be shown below, these estimates are meaningless when considering practical situations, since no characteristics of sensors resolution are taken into account. As was shown in [11], the EFPI sensor resolution is mainly determined by signal-to-noise ratio (SNR) of the registered signal $S_j(\lambda)$. SNR, in turn, is related to the amplitudes $S'_{Mj}$, considered in section III.A and to the noises, analyzed in section III.B. It also should be noted that other mechanisms, considered in section III.C, producing sensor crosstalk can arise.

## III. Theoretical analysis

### A. Amplitudes of sensors responses

In parallel configuration (see fig. 2, (a) for example) the light intensity brought to each interferometer is determined by the division ratios at the coupling elements. For the case of uniform power distribution between the sensors, the amplitudes $S'_{Mj}$ can be expressed according to (2) as

$$S'_{Mj} = 2\pi n_j w_0^2 \cdot \sqrt{R_{1j} R_{2j}} \Big/ \left[ N^2 \sqrt{L_j^2 \lambda^2 + \left(\pi n_j w_0^2\right)^2} \right], \quad (6)$$

where $n_j$ is refractive index inside the $j$-th EFPI cavity, $R_{1j}$, $R_{2j}$ are the mirrors' reflectivities in the $j$-th EFPI, and the multiplier $N^2$ arises from the light directivity in 1-by-$N$ coupler. In some cases, a nonuniform distribution of the light power over the sensors can be used, in this case the estimate (6) must be modified. The lengths of the feeding fibers must be chosen sufficiently different, with differences much greater than $\lambda_0^2/4\pi n\Delta'$ in order to suppress parasitic interference signals. For typical values $\Delta' \sim 10$ pm the order-of-meter feeding fiber differences will be applicable. Therefore, the component $H(\lambda)$ in (4) will be stipulated only by the higher-order harmonics of the Airy function ((1) contains only the first harmonic of the Airy function) and some noises, described in section B.

In serial scheme (see fig. 2, (b) for example) the interferometers are connected to the sensor interrogator one after another, therefore, considering the signal of the $j$-th interferometer, one must take into account the light propagation through the preceding interferometers. Spectrum of the light, reflected from the $j$-th EFPI and captured by photodetector, can be written as a product $[S_{T1}(\lambda)\cdot \ldots \cdot S_{Tj-1}(\lambda)]^2 \cdot S_j(\lambda)$. According to the Gaussian beam formalism [12], taking into account two beams, travelled once and trice through the EFPI cavity, $S_{Tj}(\lambda)$ can be written in analogy with the spectral function for the reflected light $S_j(\lambda)$ as follows

$$S_{Tj}(L_j,\lambda) = T_{1j} T_{2j} \{\eta(L_j/2) + R_1 R_2 \eta(3L_j/2) \\ + 2[R_{1j} R_{2j} \eta(L_j/2) \eta(3L_j/2)]^{1/2} \cos[4\pi nL_j/\lambda + \gamma_T(L_j,\lambda)]\}, (7)$$

where $T_{1,2} = 1 - R_{1,2}$, $\gamma_T(L,\lambda) = \psi(L/2) - \psi(3L/2) + \varphi_T$, $\varphi_T$ is induced by the mirrors' reflections, typically for dielectric mirrors $\varphi_T=0$. The first two summands in (7) are constant and determine the mean transmitted optical power, whereas the third (oscillating) term doesn't affect the amount of mean transmitted power. Therefore, the expression for $S'_{Mj}$ can be written in the following form

$$S'_{Mj} = 2\sqrt{R_{1j} R_{2j} \eta(L_j)} \cdot \prod_{k=1}^{j-1} \{T_{1k} T_{2k} [\eta(\tfrac{1}{2} L_k) + R_{1k} R_{2k} \eta(\tfrac{3}{2} L_k)]\}^2$$

$$\approx \frac{2\pi n_j w_0^2 \sqrt{R_{1j} R_{2j}}}{\sqrt{L_j^2 \lambda^2 + (\pi n_j w_0^2)^2}} \cdot \prod_{k=1}^{j-1} \left[ \frac{T_{1k} T_{2k} (\pi n_k w_0^2)^2}{0.25 \cdot L_k^2 \lambda^2 + (\pi n_k w_0^2)^2} \right]^2, \quad (8)$$

where in the final expression the term $R_1 R_2 \eta(3L/2)$ was neglected comparing to $\eta(L/2)$ due to typically small values of the mirrors reflectivities in case of serial scheme.

### B. Noise sources analysis

An extensive study of single EFPI displacement sensors resolution limits with wavelength-scanning interrogation was done in [11]. It was shown that the main noise sources are:

1. Absolute wavelength scale shift $\Delta\lambda_0$, determined by the triggering fluctuations of the scanning start, $\sigma_{\Delta\lambda}=\text{stdev}\{\Delta\lambda_0\}$.

2. Jitter of the wavelength points $\delta\lambda_i$, caused by the fluctuations of the signal sampling moments, $\sigma_{\delta\lambda}=\text{stdev}\{\delta\lambda\}$.

3. Additive noises $\delta s_i$, produced by the photo registering units, by the light source intensity noises, etc. $\sigma_s=\text{stdev}\{\delta s\}$.

The first mechanism provides the shift of the measured interferometer spectrum, inherently shifting the displacement sensor readings at a value $\delta L$ [11], given by:

$$\delta L \approx - \Delta\lambda_0 \cdot L_0/\lambda_0. \qquad (9)$$

This mechanism will be directly replicated in the system of multiplexed EFPI sensors.

The second and the third mechanisms produce the noises, added to the ideal interferometer spectra $S_j(\lambda)$. As a result, the spectrum approximation procedure gives a result $L_{rj}$, which is a random value with a standard deviation $\sigma_{Lrj}$. Doubled $\sigma_{Lrj}$ value is often used as a figure of merit of sensor resolution.

The jitter of spectral points during interrogation produces the distortion of the measured spectral function $S_j(\lambda)$. The resulting signal-to-noise ratio $SNR_J$ in case of a single sensor can easily be estimated by simple trigonometric derivations. The variance of the noise produced by the wavelength jitter on $j$-th spectral function can be derived from [11, eq. (9)] and spectrum amplitude $S_{Mj}$

$$\sigma_{Jj}^2 = \tfrac{1}{2} S_{Mj}^2 / SNR_J = \left(4\pi n_j L_j \sigma_{\delta\lambda} S_{Mj}/\lambda_0^2\right)^2, \qquad (10)$$

where the expressions for $S_{Mj}$ were not substituted in order to avoid the excessive bulkiness. For the case of multiplexed sensors the noises produced by all the sensors will influence each of them.

Considering the third mechanism, one has to take into account that generally the noise variance can depend on the mean optical power, incident to the photodetector (shot noise level and laser intensity noise influence are strongly related to the mean power level). The dependency can be adequately approximated by a power function [11], and explicitly expressed via the setup parameters as

$$\sigma_s = aP^b = \text{RIN}\cdot P + \text{NEP}, \qquad (11)$$

$$P = P_0 \cdot \sum_{i=1}^{N}\left(R_{1i}^* + R_{2i}^*\right), \qquad (12)$$

$P$ is the mean optical power from all interferometers, incident on the photodetector; RIN is relative intensity noise of the utilized laser, NEP is noise equivalent power of the photodetector, both RIN and NEP are recalculated to the whole photodetector frequency band; $P_0$ is optical power irradiated by the light source; $R_{1,2}^*$ are effective mirrors reflectivities, taking into account light losses due to divergence of a non-guided beam inside the cavity and in splitting elements. The second equality in eq. (11) implies that the photodetector shot noise is neglected comparing to thermal and electronic noises. The parameters $a$ and $b$ must be obtained explicitly for a given interrogating unit.

In case of parallel scheme effective reflectivities can be written as

$$R_{1j}^* = R_{1j}/N^2, \; R_{2j}^* = R_{2j}\eta(L_j)/N^2, \qquad (13)$$

as in (6), $N^2$ arises from the light directivity in 1-by-$N$ coupler.

In case of serial scheme

$$R_{1j}^* = R_{1j}\left[\prod_{i=1}^{j-1} T_{1i}T_{2i}\eta\left(\tfrac{L_i}{2}\right)\right]^2, \; R_{2j}^* = R_{2j}\eta(L_j)\left[\prod_{i=1}^{j-1} T_{1i}T_{2i}\eta\left(\tfrac{L_i}{2}\right)\right]^2. \quad(14)$$

In order to calculate the resulting noises, one needs to sum the variances of the noises produced by the individual interferometers. After that the SNR value for each sensor can be found, allowing one to estimate the lower resolution limit according to a Cramer-Rao bound [13]. As was discussed in [11], the relation of spectral function SNR and resultant OPD standard deviation $\sigma_{Lr}(SNR)$ can be approximated as

$$\sigma_{Lr}(SNR) = C \cdot SNR^{-1/2}, \qquad (15)$$

for the approximation-based cavity measurement approach [3] with number of spectral points $M=20001$ the value $C$ found during numerical simulation was $C = 1.1 \cdot 10^{-3}$ [µm], which is close to a value determined by a Cramer-Rao bound $C = 0.9 \cdot 10^{-3}$ [µm].

The expression for resultant SNR (determined by combined laser intensity, photodetector and jitter-induced noises) can be found by joining $\sigma_s^2$ and all $\sigma_{Jj}^2$ ($j=1..N$), and is written as follows

$$SNR_j = \tfrac{1}{2} S_{Mj}^2 / \left(\sigma_s^2 + \sum_{i=1}^{N}\sigma_{Ji}^2\right). \qquad (16)$$

Finally, the expression for the standard deviation of the measured OPD $L_{rj}$ can be obtained by combining expressions (9), (15) and (16) and taking into account the dispersion summation rule and will be written as follows

$$\sigma_{Lr_j} = \left\{\left[2C^2 \cdot \left(\sigma_s^2 + \sum_{i=1}^{N}\sigma_{Ji}^2\right)\right] / S_{Mj}^2 + \left(L_j/\lambda_0\right)^2 \cdot \sigma_{\Delta\lambda}^2\right\}^{1/2}. \quad(17)$$

Substituting (6) or (8), (10) and (11) to (17), one will be able to obtain the final explicit expression, which isn't done due to excessive bulkiness.

### C. Sensors crosstalk analysis

The problem of cross-talk in multiplexed interferometric sensors is quite important [14], [15], however, for EFPI it is often neglected [8]. However, for ultra-high resolution sensors its sources must properly be taken into account.

Let us consider the serial setup first. As was mentioned in section A, the oscillating summand in (7) doesn't affect the mean transmitted optical power; instead, it will modulate the light spectrum brought to the sensors after the $k$-th one. As a result, parasitic components proportional to $S_j(\lambda) \cdot S_T^2(L_k, \lambda)$, $j > k$ will arise in $S_{MULT}$. In such a manner, the overall spectrum $S_{MULT}$ will contain parasitic components of the form $S(|L_j \pm L_k|, \lambda)$ (as can be shown by simple trigonometric derivations, the product of two harmonic functions with different frequencies is equal to a sum of two harmonic functions with combinations of the initial frequencies), as well as the $S(m \cdot L_j, \lambda)$, $m$ is natural.

For parallel configuration, parasitic components in $S_{MULT}$ will have only equivalent OPDs equal to the multiples of $L_j$, expressed as $S(m \cdot L_j, \lambda)$, $m$ is natural.

The mechanisms of parasitic components formation are also illustrated in fig. 1 in terms of interference of different beams. Solid lines below the schematic illustration of the serially multiplexed EFPIs indicate the interfering beams propagation, determining the OPDs (target and parasitic). The corresponding amplitudes $S_M$ of the components are presented on the right from the optical paths.

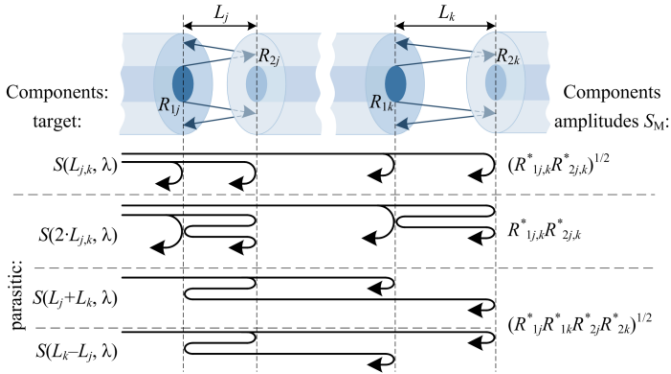

Fig. 1. Demonstration of parasitic OPDs forming.

As can be seen from the mentioning above and fig. 1, the main origin of the crosstalk in the considered systems is coincidence of the oscillation periods of the parasitic $H(\lambda)$ and the target $S_j(\lambda)$ components in (4). In order to avoid the cross-talk, the following condition on the cavity lengths must be fulfilled

$$|L_j - (p \cdot L_k + q \cdot L_l)| \gg \Delta L, \qquad (18)$$

$\Delta L$ is given by (5), $p, q$ are integer, $j,k,l=1\ldots N$, natural, and unequal to each other. The worst cases are $p=2$, $q=0$ (producing $S(2L_j, \lambda)$) or $p=\pm 1$, $q=\pm 1$, $|k-l|=1$ (producing $S(L_k \pm L_j, \lambda)$), demonstrated in fig. 1. Considering a sum of target and parasitic quasi-harmonic signals of form (1) with close frequencies and significantly different amplitudes $A$ and $a$ ($A \gg a$) we obtain $A \cdot S(L_1, \lambda) - a \cdot S(L_2, \lambda) \approx A \cdot \cos\{4\pi n[L_1 + (a/A) \cdot (L_1 - L_2)]/\lambda + \gamma\}$, producing the following error

$$\delta L \approx (a/A) \cdot (L_1 - L_2). \qquad (19)$$

Values of $A$ and $a$ can be analyzed separately, in analogy with sensors response amplitudes. For a practical case $a/A$ about 0.01 (see figs. 3 and 4 for example) and target $L_2$ deviation about 100 nm the resulting shift of the registered OPD from the real one will be about 1 nm, which is more than one order greater than the achievable displacement resolutions (see tables I and II).

As follows from the fig. 1, the level of parasitic components is strongly related to the level of target components. As can be concluded from eq. (19), the crosstalk level is determined by the ratio of parasitic and target components amplitudes, which, for the worst cases mentioned above are proportional to $(R^*_1 R^*_2)^{1/2}$. Therefore, one of the possible ways to eliminate the crosstalk influence is to use the mirrors with low reflectivities. However, this will also reduce the target component amplitude $S_M$, in turn, reducing the sensor resolution, as follows from (17).

As it can be concluded from the sections A, B and C, the parallel configuration is preferred comparing to the serial one. The first reason for this is that the influence of preceding sensors sufficiently decreases the performance of the subsequent ones. Therefore, attaining comparable performances in a serial system with relatively large number of multiplexed sensors ($N \geq 4 \div 5$) is a great challenge, requiring either beam collimating optics [16], complicating the fabrication process; or utilizing interferometers with significantly different parameters, which also makes the fabrication of the whole system more complex. The second reason is caused by the sensor crosstalk – as can be concluded from the inferences above and eq. (18), in serial scheme the number of parasitic OPDs grows as a power of two with respect to the sensors number (instead of linear dependence for the parallel scheme). Since the attempts to reduce the cross-talk influence by means of reflectivities variation will inevitably reduce the sensors resolution, a better choice will be to use EFPIs with proper OPDs, satisfying the condition given by eq. (18).

### D. Multiplexed sensors number limitation

An important characteristic is a maximal number of multiplexed sensors, which can be interrogated by a given setup, providing some determined resolution. For the cases of TDM and DWDM a great progress in this question has been made [17], whereas for the case of wavelength-domain interrogation with FDM no such analysis is still present in the literature. Such analysis is seriously complicated due to a great amount of optical scheme parameters, determining the properties of the interferometric signal (all $L_j$, $R_{1j}$, $R_{2j}$). The standard deviation of the noise-induced OPD fluctuations (see eq. (17)) is commonly used as a figure of merit of sensor resolution. In order to obtain an explicit expression, relating the attained sensors resolutions with the optical setup parameters and number of multiplexed sensors, the situation with the following assumptions, simplifying the analysis, will be considered:
- Parallel configuration is used.
- Spacing between the OPDs is uniform and equal to the smallest OPD, so that $L_j = j \cdot L_{sp}$. This implies that the sensors crosstalk is neglected, since the violation of eq. (18) in case $p=2$, $q=0$ will be produced. However, in practice, this isn't a strict limitation, since the $L_{sp}$ defines the mean spacing between the OPDs, and the crosstalk can be eliminated by relatively small offsets.
- Intra-cavity media have equal refractive indexes $n_j = n$.
- The additive noises are induced by two main sources (as in the rightmost part of eq. (11)) – laser intensity noises (determined by RIN) and photodetector noises (determined by NEP).
- Effective reflections of the mirrors, forming the EFPI cavities are equal for all the cavities: the reflectivities of the first mirrors $R_{1j}$ are all equal $R_{1j}=R_1$, the reflectivities of the second mirrors $R_{2j}$ are selected such that all effective reflectivities $R_{2j} \cdot \eta(L_j)$ are equal to some value $R_2^{**}$, therefore, the fringe visibility of the signal, produced by each sensor is equal to $V_0 = 2 \cdot (R_1 \cdot R_2^{**})^{1/2}/(R_1+R_2^{**})$.

Considering jitter-induced noises, the wavelength jitter will be transformed to additive noises by each $S_j(\lambda)$ in (4) in the following manner:

$$\sigma^2_{Jj} = \left(4\pi n \sigma_{\delta\lambda} S_M L_{sp} j / \lambda_0^2\right)^2, \quad S_M^2 = 4R_1 \cdot R_2^{**}/N^4. \qquad (20)$$

Since the practical OPD spacing can't be uniform, the spectral functions $S_j(\lambda)$ will be uncorrelated, producing uncorrelated jitter-induced noises. Therefore, the overall influence of the wavelength jitter can be found by summing $\sigma_{Jj}^2$ over all sensors:

$$\sum_{j=1}^{N}\sigma_{1j}^{2}=\left(4\pi n\,\sigma_{\delta\lambda}S_{M}L_{sp}/\lambda_{0}^{2}\right)^{2}\cdot\left(N^{3}/3+N^{2}/2+N/6\right). \quad (21)$$

For additive noises, induced by laser intensity noises and photodetector noises:

$$\sigma_{s}^{2}=\mathrm{RIN}^{2}N^{2}\left(R_{1}+R_{2}^{**}\right)^{2}+\left(\mathrm{NEP}/P_{0}\right)^{2}. \quad (22)$$

Taking into account (21) and (22), the eq. (17) is modified to

$$\sigma_{Lr_{j}}=\sqrt{2C^{2}\cdot\left[\frac{\mathrm{RIN}^{2}}{V_{0}^{2}}\cdot N^{6}+\frac{\mathrm{NEP}^{2}}{P_{0}^{2}\cdot R_{1}R_{2}^{**}}\cdot\frac{N^{4}}{4}+\left(\frac{4\pi n\,\sigma_{\delta\lambda}L_{sp}}{\lambda_{0}^{2}}\right)^{2}\cdot\left(\frac{N^{3}}{3}+\frac{N^{2}}{2}+\frac{N}{6}\right)\right]+\left(\frac{L_{sp}\cdot\sigma_{\Delta\lambda}}{\lambda_{0}}\right)^{2}\cdot j^{2}}\,. \quad (23)$$

The eq. (23) can be used for estimating the standard deviations of the measured OPD $L_{rj}$ fluctuations, induced by interrogating unit for each sensor. Since the amplitudes of the spectral functions are considered equal, the noise influence will be the same for all the interferometers, and the difference will be only in the impact of the wavelength scale shift. Therefore the lower bound of the $\sigma_{Lrj}$ estimate will correspond to the EFPI with the smallest OPD and be given by $\sigma_{Lr1}$, and the upper bound of the $\sigma_{Lrj}$ estimate will correspond to the EFPI with the greatest OPD and be given by $\sigma_{LrN}$.

Considering the asymptotic for relatively large $N$, the above-mentioned mechanisms can be sorted in a decreasing order of impact on the $\sigma_{LrN}$ in the following order: laser intensity noise, photodetector noise, wavelength jitter and wavelength scale shift.

Taking the power of two of eq. (23), setting $j=N$, specifying the maximal admissible $\sigma_{Lrj}$ value and solving eq. (23) with respect to $N$, one can obtain an estimate of maximal number of sensors that can be interrogated with the predefined resolution.

It also should be noted that the considered situation can be generalized to the case of serial scheme. However, much more complicated rules on the relation of sensors' OPDs and mirrors' reflectivities must be fulfilled in order to meet the assumptions, made in the last point above, implying the requirements on the mirrors reflectivities.

Let us mention two points essential for understanding the performance limitations of the multiplexed sensors:

1. The noises formed by all interferometers influence the SNR of all the interferometric signals; hence, each interferometer affects the resolution of all the others.

2. When using OPD demodulation approaches close to optimal [3], [18], the filtering performed in order to extract the partial spectra $S_{j}(\lambda)$ does not need to be taken into account in Eqs. (16), (17) and (23) in case of white noises (which is a typical situation). This is so because of the equivalent filtration with resolution $\Delta L$, performed by the approximation procedure (this filtration is also taken into account in the derivation of a Cramer-Rao bound).

## IV. EXPERIMENTAL DEMONSTRATION

The both serial and parallel schemes were implemented in the experiment. Spectra measurements were performed using the optical sensor interrogator NI PXIe 4844, utilizing a tunable laser with SMF-28 single-mode fiber output and the following parameters: scanning range [1510; 1590] nm; wavelength step $\Delta=4$ pm; spectral points number $M=20001$; photodetector frequency band 570 kHz; wavelength jitter stdev $\sigma_{\delta\lambda}=1$ pm; wavelength scale shift stdev $\sigma_{\Delta\lambda}\approx 0.05$ pm; $a=8.47\cdot10^{-4}$ and $b=0.81$ for eq. (11), with the corresponding RIN $\approx 3\cdot10^{-4}$ in the full frequency band, RIN $\approx -130$ dB/Hz, NEP $\approx 3.5\cdot10^{-11}$ W in the full frequency band, NEP $\approx 1.1\cdot10^{-13}$ W/$\sqrt{\mathrm{Hz}}$; output optical power $P_{0}\approx 0.06$ mW; spectra acquisition rate about 1 Hz. Signal processing was running on a PXI chassis controller NI PXIe 8106.

In order to simplify the processing (avoid the peak tracking in the Fourier transform (FT) representation of $S_{\mathrm{mult}}(\lambda)$), we assumed that the maximal OPD deviations are less than 30 μm. Therefore, the partial spectra $S_{j}(\lambda)$ could be extracted by pre-defined band-pass filters, which was done by the same way as in [8]. After that, to each $S_{j}(\lambda)$ spectrum the approximation-based approach described in [3] was applied.

### A. Serial scheme

The realized serial multiplexing scheme is shown in fig. 2 (b). The parameters of the optical setup were the following: interferometers OPDs $L_{1}=42$ μm, $L_{2}=170$ μm, $L_{3}=250$ μm (in the second cavity a mirror with dielectric multilayer evaporation, producing an increased reflectance $R_{2}\approx 20\%$ was used, the third cavity was formed by a plate of a crystalline silicon with $R_{1}\approx 16\%$, $R_{2}\approx 31\%$ [19], all other $R=3.5\%$ were stipulated by Fresnel reflections at air-glass bound).

In fig. 3 the measured spectral function $S_{\mathrm{mult}}(\lambda)$ (a) and its FT (b) are shown. In fig. 3 (b) the abscissa axis corresponds to the air-gap cavity length, by normalizing the FT domain $x$ axis on $\lambda_{0}^{2}/4\pi n$. For the Si-plate EFPI it should be properly recalculated with taking into account Si refractive index with dispersion. It is clearly seen in fig. 3 (b) that in the spectral function of serially multiplexed EFPIs, parasitic components corresponding to OPDs combinations ($L'_{3}\pm L_{2}$) are presented along with single, double and triple OPDs.

The interferometers OPDs $L_{rj}$ were demodulated from the registered spectra as mentioned above. The standard deviations of the estimated OPDs $\sigma_{Lrj}$, shown in table I (second column), were calculated over the temporal intervals about 10 minutes, corresponding to 600 OPD samples for each sensor. The analytical estimations of the standard deviations calculated according to (17) with substituted parameters corresponding to the ones of the experimental setup are also presented in the third column for reference, demonstrating good correspondence with the experimental values. In order to compare the resolution decrease in multiplexed system versus single-sensor system, the OPD standard deviations of the single sensors with the OPDs corresponding to the ones of the multiplexed sensors are shown in the fourth column (these experiments are present in [11], [19]).

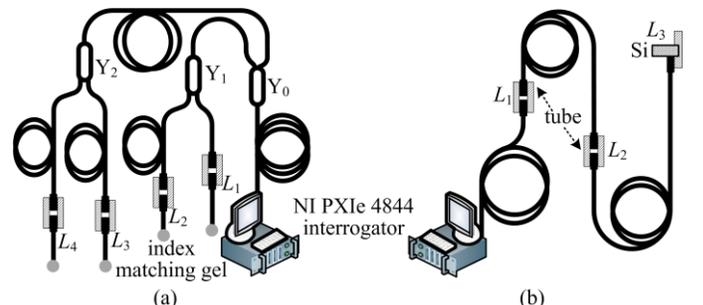

Fig. 2. Experimental setups for parallel (a) and serial (b) configurations.

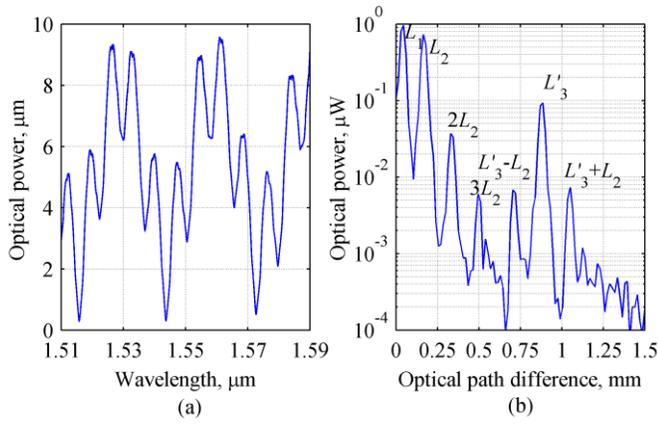

Fig. 3. Serial scheme with 3 EFPIs: spectral function (a) and its FT (b).

TABLE I
Measured OPD standard deviations, serial scheme

| EFPI OPD, µm | $\sigma_{Lr}$, pm | | Single, experiment |
|---|---|---|---|
| | Multiplexed | | |
| | Experiment | Estimated, (17) | |
| 42 | 32 | 28 | 8, [11] |
| 170 | 43 | 39 | 10, [11] |
| 250 | 78 | 71 | 10, [19] |

*B. Parallel scheme*

The realized parallel multiplexing scheme is shown in fig. 2 (a). The parameters of the optical setup were the following: interferometers OPDs $L_1$=41µm, $L_2$=195µm, $L_3$=526µm, $L_4$=719µm; all mirrors reflectivities were equal to $R$=3.5% (stipulated by Fresnel reflections at the air-glass bound).

In fig. 4 the measured spectral function $S_{mult}(\lambda)$ (a) and its FT (b) are shown, again in fig. 4 (b) the abscissa axis is renormalized to the air-gap cavity length domain. It should be noted that parasitic components with OPDs combinations are absent in the parallel scheme. The $S_{mult}(\lambda)$ was processed as mentioned above, standard deviations of the measured $L_{rj}$ values are presented in table II.

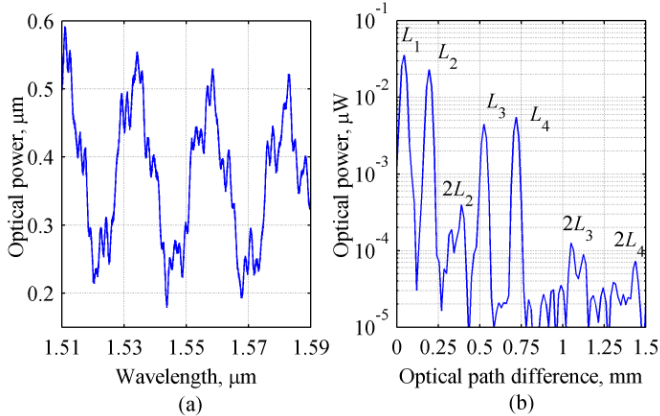

Fig. 4. Parallel scheme with 4 EFPIs: spectral function (a) and its FT (b).

As can be seen from the second column of the table II, the initially implemented configuration with equal mirrors reflectivities is not optimized, since the standard deviations of different sensors are sufficiently unequal. Therefore, a more accurately developed distribution of the reflectivities was used, providing higher fringe visibilities for all interferometric signals $S_j(\lambda)$. The interferometer OPDs and $R_1$ were not modified, but the reflectivities of the second mirrors were changed to the following: for $L_1$=41µm – $R_2$=3.5%; for $L_2$=195µm – $R_2$=20%; for $L_3$=526µm and $L_4$=719µm – $R_2$=90%. The standard deviations obtained with an optimized setup are presented in fourth column of the table II. It should be noted that the stdev estimations made according to (17) with substituted parameters corresponding to the ones of the experimental setup are in a good accordance with the experimental observations. The discrepancies can be explained by nonuniform power splitting in Y-couplers. Also it should be noted that the resolutions of the multiplexed sensors in optimized configurations are mostly 2-5 times worse than for single sensors with the same OPDs.

By the example of parallel setup the relation of maximal number of multiplexed sensors with the attained standard deviation of the sensors' readings was studied and is presented in fig. 5.

The experimentally attained standard deviations of the measured OPDs for single sensor (the results from [11] are presented), two sensors (multiplexed with a Y-coupler in parallel, $L_1$=120µm, $L_2$=350µm, $R_{11}$=$R_{12}$=3.5%, $R_{21}$=20%, $R_{22}$=90%) and four sensors in optimized configuration are compared with the analytical estimation made according to (23) with the following parameters: $L_{sp}$=180µm, $R_1$=$R_2^{**}$=3.5%, the rest parameters the same as for the utilized in the experiments interrogating unit (see the beginning of the current section). The parameters $L_{sp}$, $R_1$ and $R_2^{**}$ were selected maximally close to the experimental ones, averaged over all the interferometers. The analytical curves were calculated for the two cases: the 1-st sensor (lower bound of the attained resolution) and the $N$-th sensor (upper bound of the attained resolution). Considerable discrepancies between the theoretical predictions and experimental results for the 3-rd and the 4-th sensors in the 4-sensor setup are due to lower values $R_2^{**}$ for these interferometers than was substituted into the eq. (23).

TABLE II
Measured OPD standard deviations, parallel scheme

| EFPI OPD, µm | $\sigma_{Lr}$, pm | | | | single, exp.,[11] |
|---|---|---|---|---|---|
| | Nonoptimized | | Optimized | | |
| | Exp. | Est., (17) | Exp. | Est., (17) | |
| 41 | 41 | 28 | 42 | 33 | 8 |
| 195 | 63 | 74 | 37 | 37 | 14 |
| 526 | 180 | 190 | 76 | 48 | 23 |
| 719 | 200 | 258 | 84 | 65 | 34 |

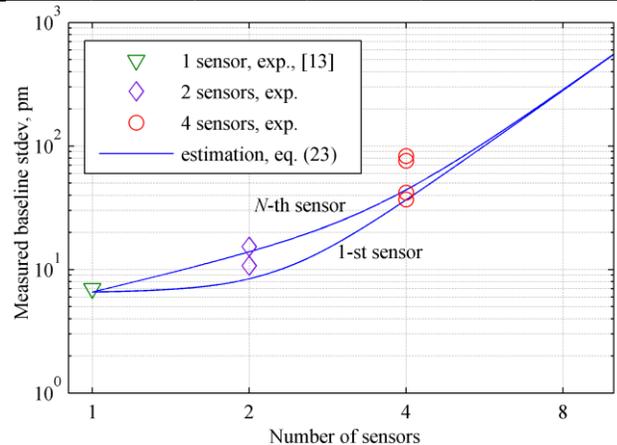

Fig. 5. Sensors resolution dependence on the number of sensors.

### C. Sensors crosstalk considerations

In order to study the sensing abilities of the assembled system and ensure the cross-talk absence, we have consequently heated and cooled one of the sensors, whereas the rest were placed into the thermally isolated chamber. This operation was repeated for the both configurations and all combinations of heated/isolated sensors. In fig. 6 the corresponding OPD variations curves are illustrated. In fig. 6 (a) the sensors readings during the heating of the 1-st sensor in serial scheme is illustrated, in fig. 6 (b) – the heating of the 4-th sensor in parallel scheme (non-optimized). For illustrative purpose the scales of the non-perturbed sensors OPDs were 10 times magnified, and additional initial shifts were subtracted from the curves. OPD deviations of the non-perturbed sensors demonstrate no correlation with the perturbed ones, hence, no cross-talk was present in the current systems, at least, at the levels limited by the attained sensors resolutions. The range of the temperature changes was around 15 K, on this basis the air-gap sensors' OPD temperature sensitivity can be estimated about 6.4 nm/K.

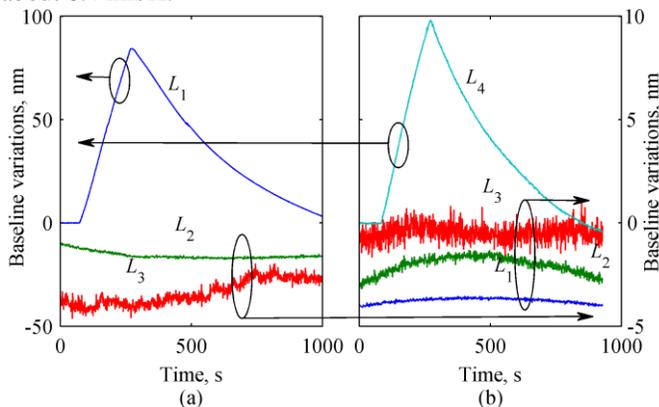

Fig. 6. Demonstration of temperature sensitivity of the multiplexed sensors for serial (a) and parallel (b) configurations.

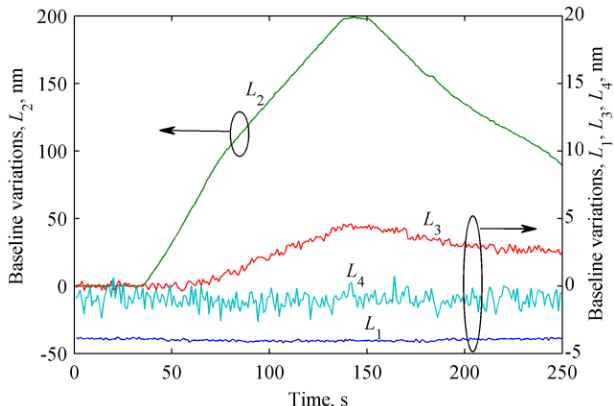

Fig. 7. Crosstalk demonstration in a deliberately modified parallel scheme.

In a separate experiment with parallel scheme the value $L_3$ was changed to ~390 μm to match the doubled $L_2$. In such a manner, by heating the 2-nd sensor the cross-talk effect ($L_2$ changes affecting the 3-rd sensor readings) was demonstrated. In fig. 7 the corresponding signals are shown, the additional shifts were subtracted from the curves for better demonstrativeness. For the same reason the scales for $L_1$, $L_3$ and $L_4$ variations were 10 times magnified. The level of the parasitic deviation of $L_3$ is in accordance with the expression (19).

## V. CONCLUSION

The vulnerability of the multiplexed EFPI sensors to parasitic crosstalk effects has been demonstrated. As shown by analytical and experimental investigations, for a given number $N$ of multiplexed sensors, the serial configuration is more subject to the cross-talk than the parallel one. With the use of the developed mathematical model one is able to estimate the achievable sensors resolutions for a given optical setup and interrogating unit. In a general case of large sensors number and arbitrarily different interferometers, the inverse task (for example, finding the maximum number of sensors that can be interrogated by a given device with some defined resolution) can't be solved analytically. Nevertheless, in case of equal products and sums of the effective mirrors' reflectivities $R_{1j}^* \cdot R_{2j}^*$, $R_{1j}^* + R_{2j}^*$ for all $j=1...N$, an explicit equation, relating the attainable resolution with the sensor number was derived.

The sensors' OPDs must be chosen such that: enough spacing between the OPDs is left in order to perform frequency-division demultiplexing without cross-talk; parasitic equivalent OPDs (caused by light propagation in the EFPI cavity and cross-influence of the sensors in serial scheme) do not coincide with the main OPDs $L_j$, violating eq. (18); fringe visibilities of the interference signals produced by all the interferometers are maximally close to unity and spectrum amplitudes $S_{Mj}$ are maximally close to each other.

**Nikolai A. Ushakov** received the M.S. degree in radiophysics in 2011 from St. Petersburg Polytechnic University, Russia. He is currently working towards the Ph.D. degree on interferometric fiber optic sensors in St. Petersburg Polytechnic University.

His scientific interests include interferometric measurement techniques, optical microcavities and distributed optical fiber sensors.

**Leonid B. Liokumovich** received the M.S. degree in radiophysics in 1992, candidate and doctor of physic-mathematical science degrees in 1995 and 2008 from St. Petersburg Polytechnic University, Russia. He is a professor at St. Petersburg Polytechnic University, Russia.

His research interests include single- and multimode fiber-optic interferometric systems, phase and polarization sensors, processing of interferometric signals.